\documentclass{article}
\usepackage{graphicx} 

\usepackage{caption,subcaption,graphicx}
\usepackage{amsmath,amsfonts,amssymb}
\usepackage{graphicx}
\usepackage[misc]{ifsym}
\usepackage[colorlinks=true, allcolors=blue]{hyperref}
\usepackage[margin=2.5cm]{geometry}
\def\BibTeX{{\rm B\kern-.05em{\sc i\kern-.025em b}\kern-.08em
    T\kern-.1667em\lower.7ex\hbox{E}\kern-.125emX}}

\usepackage{verbatim}
\immediate\write18{texcount -tex -sum  \jobname.tex > \jobname.wordcount.tex}

\providecommand{\keywords}[1]
{
  \small	
  \textbf{\textit{Keywords---}} #1
}

\author{Emile Prêle$^{a,b, \,\text{\Letter}}$, Martino Calvo$^{a,b}$, Usasi Chowdhury$^{a}$, Gregory Garde$^{a,b}$,\\ Florence Levy-Bertrand$^{a,b}$, Alessandro Monfardini$^{a,b,\,\text{\Letter}}$ \\\\
        \small $^{a}$Univ.Grenoble Alpes, CNRS, Grenoble INP, Institut Néel, Grenoble, France \\
        \small $^{b}$Groupement d'Interêt Scientifique KID, Grenoble, Saint-Martin d'Hères, France
}
\title{On-chip filled array hyperspectral kinetic inductance detectors (FHYKID) for line intensity mapping}

\date{}
\begin{document} 
\maketitle

\vspace{-1cm}

\begin{center}
\textit{This manuscript is a draft of the published proceeding paper : On-chip filled array hyperspectral kinetic inductance detectors (FHYKID) for line intensity mapping \cite{FHYKID_first_paper}.}
\end{center}

\let\thefootnote\relax\footnotetext{Further author information: \\emile.prele@neel.cnrs.fr\\alessandro.monfardini@neel.cnrs.fr}

\begin{abstract}
 The next generation
 Line Intensity Mapping (LIM) experiments will require very high throughput instrumentation, and as a consequence high pixel counts focal planes. For the study of the [CII] line in the redshift range z=7-8, we develop a hyperspectral filled focal plane (FHYKID) operating in the range 210-235 GHz and with a spectral resolution reaching 120. The spectral filtering is achieved thanks to planar Slot Split Ring Resonators (SSRR) coupled to Lumped Element Kinetic Inductance Detectors (LEKID). This robust concept achieves, thanks to its simplicity, large quantum efficiency and is scalable, even with current readout electronics systems, to several tens of thousands beams on Sky. Our FHYKID will be one of the four channels of the NIKONA instrument, currently seeking funding and to be installed at the IRAM 30-meter telescope at Pico Veleta. Among the science goals, simultaneously mapping the local (galactic and nearby-galaxies) CO line and the cosmological [CII].
\end{abstract}

\keywords{Millimeter wave, Line Intensity Mapping, Kinetic Inductance Detectors, On-chip Spectroscopy, Galactic science, Instrumentation, Superconductivity}

\section{INTRODUCTION}
\label{sec:intro}
Millimeter wave spectroscopy is currently well developed at very high spectral resolution, with several telescopes and instruments as ALMA, NOEMA, EMIR\cite{EMIR}. However, high spectral resolution instruments are limited to narrow field of view and consequently, show slow mapping speed. This trade off motivates the development for high angular resolution, fast mapping speed instruments, with intermediate spectral resolution $R=\frac{f_0}{\Delta f}=100-1000$ to quickly identify astrophysically relevant targets, conduct statistic studies and wide surveys of well defined gas emissions lines, as the CO and CII lines. We previously developed such detector arrays with first a proposition with the HYPKID design \cite{Chowdhury_HYPKID}, that provided on-chip spectroscopy with a resolution reaching $R=1000$ and high quantum efficiency. However, the current HYPKID design relies on horn antenna, which limits the pixel density on the array due to the physical size of the horns.
In this study, we developed a new kind of Filled Hyperspectral array also relying on Kinetic Inductance Detectors (FHYKID), and using Meta-materials inspired Frequency Selective Surface (FSS) to replace the HYPKID horns and filters. An array of more than 1000 pixels of 1.8 mm pitch has been designed, fabricated and tested, which allows to match any millimeter wave telescope beam size. This design has been tested and validated experimentally in lab using an optical dilution fridge. We optimized this design around the 210 - 235 GHz frequency range to allow for [CII] Line Intensity Mapping (LIM) for high redshift galaxies at z=7-8 redshift, as well as CO(2-1) LIM to study gas composition of nearby galaxies. This new detector array is the first iteration of a new type of detectors that should equip one of the four channels of the future NIKONA instrument in IRAM 30m-Telescope. 

\section{DESIGN}

\begin{figure}[h!]
    \begin{center}
    \includegraphics[width = \linewidth, page = 2]{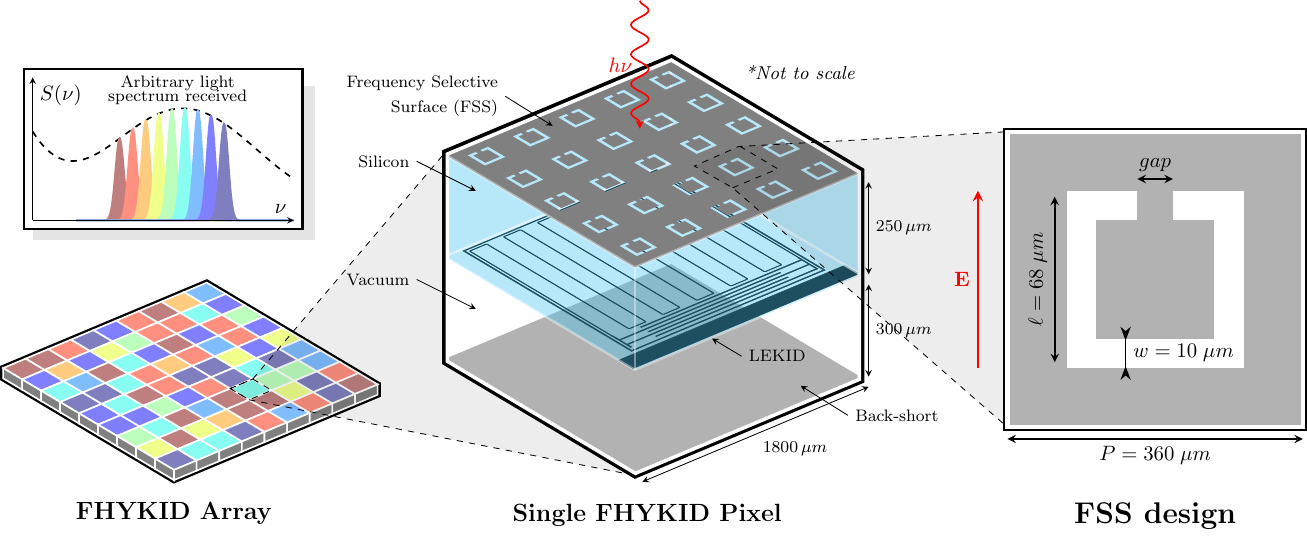}
    \end{center}
    \caption[] 
    { \label{fig:design} 
    FHYKID schematic design and principle : from the whole array to the Frequency Selective Surface.}
\end{figure}

Standard Aluminum LEKID usually developed for millimeter wave astronomy (as for the NIKA2 instrument\cite{nika2_instrument}) are known to absorb over a large photon frequency bandwidth of at least 50 GHz (spectral resolution of 3-5), that needs to be reduced to a narrower band to allow spectroscopy measurements. The idea is to create an array of randomly distributed pixels that only absorb a narrow  and tunable millimeter wave band, each with a different central frequency of absorption. An hyperspectral map of the sky (position $\times$ position $\times$ frequency) can then be constructed by sweeping the telescope beam on sky to allow each pixel to map a same sky patch (fig.\ref{fig:design}).

\begin{figure}[h!]
    \begin{center}
    \includegraphics[width = \linewidth, page = 3]{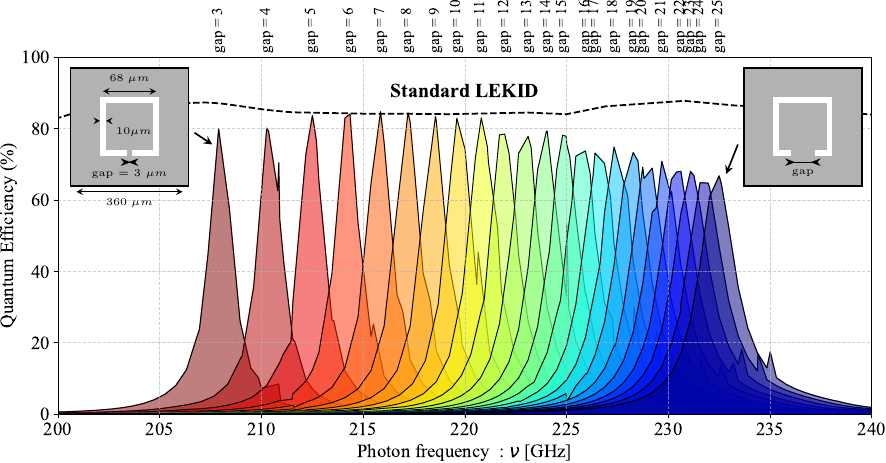}
    \end{center}
    \caption[example] 
    { \label{fig:CST_simu} 
    CST Studio simulation results of a single FHYKID pixel quantum efficiency over the photon frequency for different FSS central frequency (different gap size, in $\mu m$). Simulations of the same geometry, without FSS (standard LEKID) is also plotted in black dashed line for reference.}
\end{figure}

In the HYPKID\cite{Chowdhury_HYPKID}, the incoming free-traveling millimeter wave is coupled to a single superconducting split ring resonator (SRR), via a circular horn. The SRR resonates at a very specific frequency, and excites a LEKID inductance close-by, which allows for standard LEKID measurement. The new concept for FHYKID is to use a filtering passive metallic surface to let pass only a specific wave frequency $\nu$ to be absorbed by a standard LEKID. The frequency selective surface (FSS) must be designed to create a selective band-pass filter. Thus, the standard microstrip LEKID is illuminated through a Silicon substrate, with the front side (facing the incoming wave) being the frequency selective surface, also acting as a ground-plane. The LEKID is optimized to absorb more than 80\% of the incoming light around the 200-250 GHz bandwidth. A metallic back-short is also added to maximize the KID absorption around 220 GHz ($3\lambda/4 = 300\;\mu m$). The LEKID pitch size as been chosen to be 1.8 mm for this study which corresponds to the optimized pixel size for NIKA2 instrument on IRAM 30m Telescope around 220 GHz. However, any standard dimension could be chosen to match the pixel size with any other beam size. Compared to the HYPKID design, this new array is not limited by the horn size, and the entire image plane is filled with pixels (filled array). This allows for better sensitivity, higher scanning speed and better angular resolution.
This new detector follows a simple and robust fabrication technique, using a single 100mm silicon wafer of 250 $\mu m$ thickness, and one photo-lithography on each side of the wafer. The LEKID array is made out of superconducting Aluminum ($T_c\approx1.3\,K$, $f_{gap}\approx 100\,GHz$), to absorb the incoming photons in the 210-235 GHz band. The standard LEKID design has been simulated with \textit{Sonnet} and \textit{CST Studio} multiple times without the FSS, and tested in lab. The LEKID readout frequency is distributed randomly on the array to reduce possible cross-talk between resonators, and the coupling quality factor has been optimized to match the internal quality factor currently observed on the sky with NIKA2 ($Q_c=Q_i\approx25\times10^3$). The FSS side is made out of Niobium ($T_c\approx9.2\,K$, $f_{gap}\approx 700\,GHz$) to stay superconducting even while receiving 250 GHz photons. The FSS needs to be superconductor to prevent any loss in the metallic plane, and allow for highest quality factor filtering.
Several FSS design exists to create tunable passive band pass filters (Loaded square, Loaded circle, Jerusalem Cross) \cite{FSS_review}, however, most of them are not selective enough (R$<$50), with reasonable geometrical constrains, to be used in this application. The current design was inspired by meta-material papers using split ring resonators (SRR) as highly selective rejection filters\cite{SRR_1, SRR_2}. Therefore, the proposed design is composed of 5$\times$5 Slot Split Ring Resonators (band-pass filter) of 360 $\mu m$ pitch per FHYKID pixel. The SSRR geometry stay consistent over the whole array, except for the gap size which controls the selected central frequency, and allows to change the frequency received between different pixels. The FSS central frequency is also randomly distributed on the array to prevent crosstalk between the FSS 5$\times$5 blocks. \textit{CST Studio} simulations of the FSS show that increasing the pitch of a single FSS element allows for higher selectivity of the filter, which means higher spectral resolution $R = \Delta f/f_0$, with $f_0$ the band central frequency and $\Delta f$ the full width at half maximum. However, to prevent higher modes propagation, the pitch must be smaller than the minimum wavelength in the substrate material. Here, the maximum frequency is around 250 GHz, which limits the pitch to around 350 $\mu m$, and limits the frequency resolution, for this geometry. Another important point, is that the SSRR design is polarization sensitive, which allows only for the Inline polarization to reach the LEKID. The FHYKID geometry was validated with multiple \textit{CST Studio} simulations (fig.\ref{fig:CST_simu}) which show a clear narrowing of the received bandwidth (BW) between a standard LEKID (black dashed line, $BW=100\,GHz$) with the same geometry, and a FHYKID pixel (filled color plot, $BW=2.5-3\;GHz$). 

\begin{figure}[tb]
\centering
\begin{subfigure}[h]{0.3\textwidth}
    \centering
    \includegraphics[width=\textwidth]{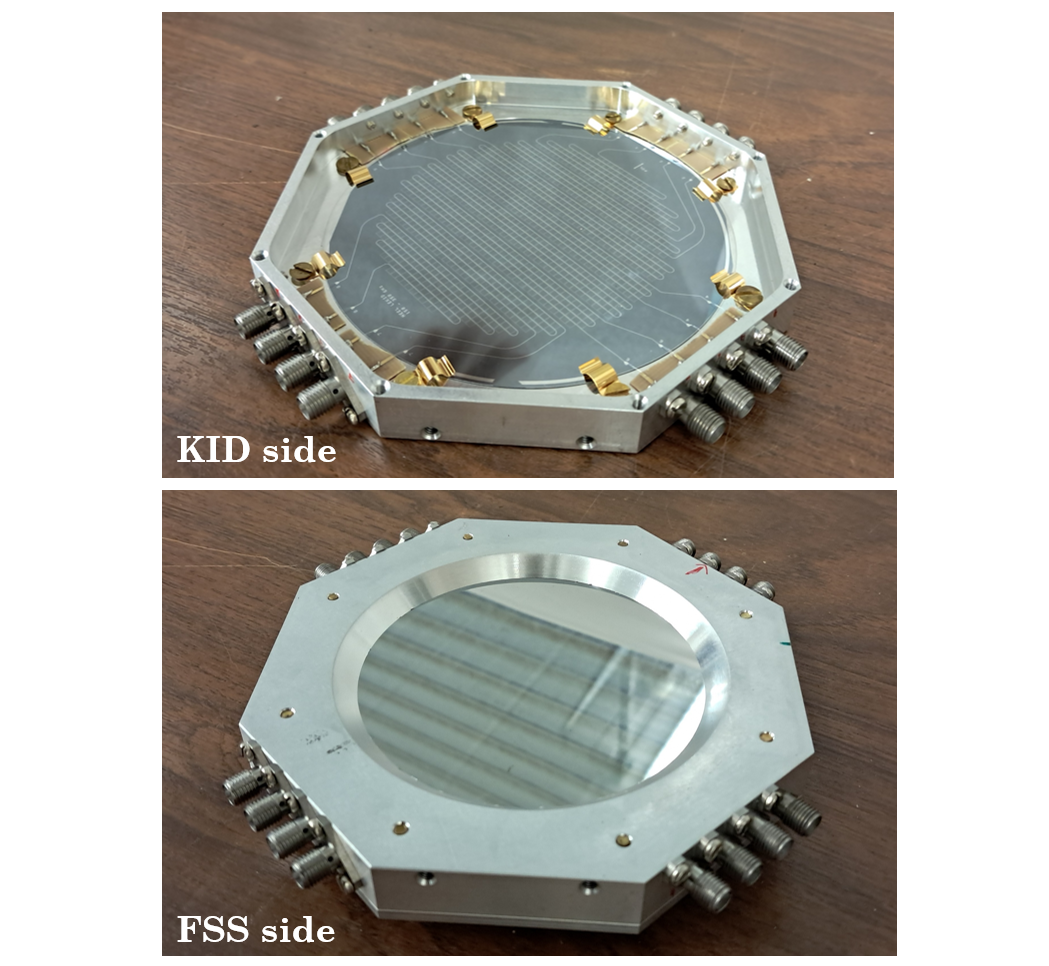}
    \caption{1128 pixels sensor assembled in the device holder} 
    \label{fig:1}
\end{subfigure}
\hfill
\begin{subfigure}[h]{0.3\textwidth}
    \centering
    \includegraphics[width=\textwidth, page = 5]{fig.pdf}
    \caption{Frequency Selective Surface seen under optical Microscope}
    \label{fig:2}
\end{subfigure}
\hfill
\begin{subfigure}[h]{0.3\textwidth}
    \centering
    \includegraphics[width=\textwidth, page = 6]{fig.pdf}
    \caption{LEKID seen under optical Microscope}
    \label{fig:3}
\end{subfigure}
\caption{Photographs of the fabricated detector}
\label{fig:fhykid_V0}
\end{figure}

The frequency resolution reached in simulation is in the range $R=80-140$, decreasing when the gap size increases (the central square metallic patch is less and less a "lumped" element). The FHYKID reaches more than 80\% of quantum efficiency, which is mainly limited by the standard LEKID absorption. Tuning the gap between $3 \;\mu m$ and $25 \; \mu m$ allows to cover from 208~GHz to 233~GHz, with a frequency resolution limited by the optical photo-lithography resolution on the gap ($1 \;\mu m$). A wider band could be covered by changing a second parameter of the design over the whole array (for example the $68\; \mu m$ side length of the square). Moreover, high off-band rejection is achieved in simulation thanks to the high order band-pass filtering effect of the FSS.  The simulated design was fabricated on a 100 mm silicon wafer, creating a 1128 FHYKID pixels array, distributed on 8 readout feedlines. The KID side is fabricated using the standard in-house process\cite{Goupy_fab_KID}, on a 25 nm e-gun evaporated Aluminum film, using mask photo-lithography development and wet chemical etching. The FSS side is done on a 100 nm Niobium layer, also using mask photo-lithography development and Reactive-ion etching. Both sides of the wafers are cleaned with Argon ion milling prior to metal deposition, to remove the semiconductor native oxide. The fabricated sensor is then mounted and bounded to the device holder (fig.\ref{fig:1}). Both side lithography are aligned using the wafer flat as a reference, because of lack of alignment marks on the LEKID mask. Therefore, a misalignment error of around 0.5~mm is to be expected on this first design. Optical microscopy images of the device show the clean results of the fabrication process, with an uv-lithography resolution reaching 1~$\mu m$ for both sides of the wafer (fig.\ref{fig:2} and \ref{fig:3}).

\newpage

\section{EXPERIMENTAL RESULTS}
\begin{figure}[h!]
\begin{center}
\includegraphics[width = 0.9\linewidth]{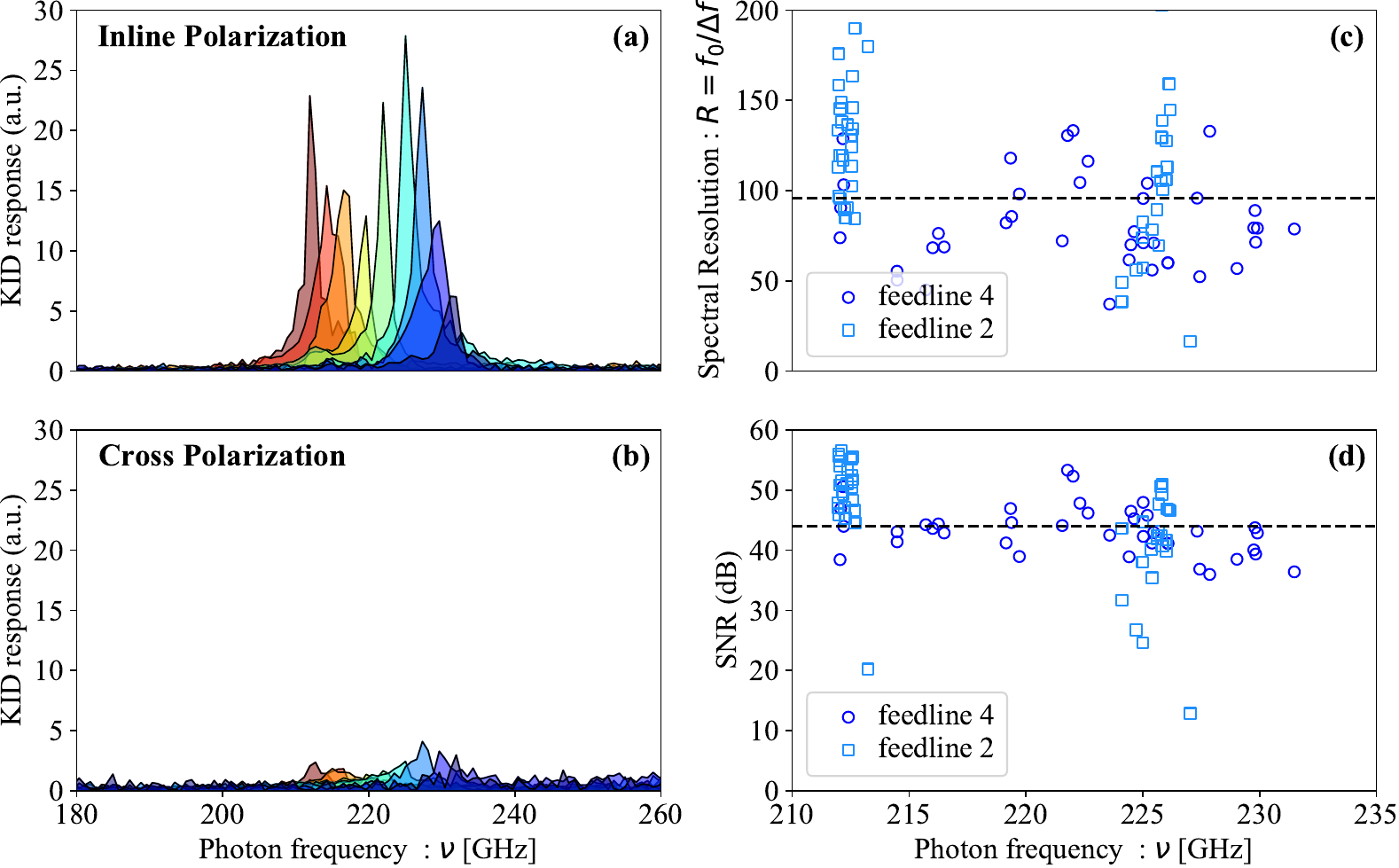}
\end{center}
\caption{Experimental characterization of the fabricated FHYKID sensor. Spectral KID response versus photon frequency for inline (a) and cross (b) polarization illumination with the MPI. Computed spectral resolution (c) and Signal to Noise Ratio (d) of the best optically responding pixels.}
{ \label{fig:results}}
\end{figure}

The fabricated FHYKID array was tested in a custom made dilution fridge with an optical window, cooling the device down to 100 mK \cite{nika2_instrument}. A Martin-Puplett Interferometer (MPI) \cite{durand:MPI} was used to measure the absorption Spectrum of the detectors (fig.\ref{fig:results}). The frequency shift of the LEKID is measured using a dedicated multiplexing electronics \cite{Bourrion_2013}, synchronized with the Martin-Puplett Interferometer data. The Inline Polarization spectrum of a few selected pixels is plotted in figure \ref{fig:results} (a.). As expected from the design, different pixels respond to different central frequency, spanning between 207 GHz and 230 GHz, with a narrow bandwidth of about 2 GHz.

\newpage

High off-band rejection is achieved thanks to the FHYKID design and no parasitic signal is observed below 200~GHz and above 240~GHz (a global 350~GHz low pass filter is installed in the dilution fridge optical path). The measure with an orthogonal polarization is also plotted fig.\ref{fig:results}(b), and shows less than 20~dB cross-polarization purity, as expected from the simulations. A fraction of the signal appearing in the Cross-Polarization measures can be due to a misalignment tilt between the two lithography layers on the detector, and/or by a tilt misalignment between the Martin-Puplett Interferometer and the sensor (meaning the "orthogonal" and "inline" measurements are probably not fully "inline" and "orthogonal"). Only two of the height feedlines were measurable during one cool-down of the cryostat due to the number of cold amplifier available on the fridge. We measured the pixels response from feedline 4 (FSS distributed randomly) and from feedline 2 (three FSS frequency block distribution) on a 500~MHz readout electronic bandwidth. Over the two measured feedlines, only a certain amount of pixels were considered as "optically responding" to the interferometer signal. Indeed, because of the two-lithography misalignment, some pixels respond with two absorption line (they face a mix of FSS geometry), and some of them do not respond at all to the optical signal (facing only metal, without SSRR). This gives in total 50 "good" pixels for feedline 2, over the 81 expected ones (62\% yield), and 39 "good" pixels for feedline 4 (expected 88 pixels, 44\% yield). A better yield is to be expected in the future design with a proper alignment of the two lithography. The spectral resolution of the "good" optically responding pixels across feedlines 2 and 4 is also plotted on figure \ref{fig:results} (c.), and shows a mean frequency resolution of about $R=95$. The spectral resolution was computed by fitting a Lorentzian curve to each response, and estimated using $R = f_0/\Delta f$. The currently tested array has no back-short, which limits the sensitivity of the detector to around 50\% of the theoretically reachable efficiency, based on \textit{CST Studio} Simulations, and therefore limits also the spectral resolution. A large disparity of the spectral resolution, going from $R=10$ to $R = 190$, is probably due to the misalignment of the two lithography. Finally, the signal to noise ratio (SNR) is also computed as $20\log(A_0/N_{rms})$ with $A_0$ the amplitude of the Lorentzian fit at $f_0$, and $N_{rms}$ the measured noise RMS value between 250~GHz and 1000~GHz, for the same pixels as the ones in plot (c). The vast majority of pixels have a SNR higher than 40~dB, which guarantees a good detection of the signal. This high SNR also proves the high off-band rejection of the background over the whole band (below 200 GHz and from 240 GHz to 1 THz). Further measurements are planned to accurately characterize the sensitivity and Noise Equivalent Power (NEP) of those detectors, using a home-made sky simulator \cite{sky-simulator} as a thermal calibrated source, to measure the response for different thermal power.

\section{CONCLUSION}
Results presented in this paper demonstrate the feasibility of polarized filled array on-chip spectrometers such as the FHYKID, with a 1128 pixels array made and tested experimentally, providing spectral resolution up to R=180, while keeping an optimal angular resolution. Such device could be easily mounted in a dilution fridge having an optical window, as NIKA2 instrument\cite{nika2_instrument}, to upgrade the standard photometric measurement and to allow for quick LIM survey for statistics measurements, or to spot interesting target to be observed with high resolution interferometers, or heterodyne instruments. The polarization selectivity of the current device allows for spectroscopy $\times$ polarimetric measurement of specific target. However, most of the standard LIM measurements are done in both polarization, so we are working on a new FSS design to allow high frequency selectivity, while allowing both polarizations to reach a Hilbert pattern shaped LEKID\cite{Hilbert_LEKID}. Further measurements will be done to accurately characterize the sensitivity of such detectors and compare it to standard LEKID. We are currently working on a new design for the 3~mm Atmospheric Window ($\approx90\;GHz$), with increased spectral resolution, up to $R=500$, for the CO(1-0) line mapping of galaxies. 

\section*{ACKNOWLEDGMENTS} 
We acknowledge the overall support of the Cryogenics, Electronics and Nanofab groups at Institut Néel and LPSC. The fabrication of the device described in this paper was conducted at the PTA Grenoble micro-fabrication facility. This work has been funded by the LabEx FOCUS through grant $n^{\circ}$ANR-11-LABX-0013 and the CNES $R\And T$ program.

\bibliography{report} 

\end{document}